\documentclass[]{spie}  
\usepackage[]{graphicx}

\title{Extinction correction and on-sky calibration of SCUBA-2} 

\author{Jessica T. Dempsey\supit{a}, Per Friberg\supit{a}, Tim Jenness\supit{a}, Dan Bintley\supit{a} and Wayne S. Holland\supit{b,}\supit{c}
\skiplinehalf
\supit{a}Joint Astronomy Centre, 660 N. A'ohoku Place, Hilo, U.S.A.;\\
\supit{b}UK Astronomy Technology Ctr, Royal Observatory, Edinburgh, U.K.;\\
\supit{c}Institute for Astronomy, University of Edinburgh, Royal Observatory, Edinburgh, UK;\\
}

\authorinfo{Further author information: (Send correspondence to J.T. Dempsey)\\JTD.: E-mail: j.dempsey@jach.hawaii.edu, Telephone: 1 808 969 6566}

\begin{document} 
  \maketitle 

\begin{abstract}
Commissioning of SCUBA-2 included a program of skydips and observations of calibration sources intended to be folded into regular observing as standard methods of source flux calibration and to monitor the atmospheric opacity and stability. During commissioning, it was found that these methods could also be utilised to characterise the fundamental instrument response to sky noise and astronomical signals. Novel techniques for analysing on-sky performance and atmospheric conditions are presented, along with results from the calibration observations and skydips.

\end{abstract}

\section{INTRODUCTION}
\label{sec:intro} 

SCUBA-2 is a multi-wavelength submillimeter bolometric array receiver which is currently being commissioned on the James Clerk Maxwell Telescope (JCMT) at Mauna Kea, Hawaii. For full details of the instrument design, performance and commissioning details see the papers by Holland et al. \cite{holland2}($\#$7741-4), and Bintley et al.\cite{bintley} ($\#$7741-5) in these proceedings. \\

Calibration of astronomical observations in the submillimeter requires an accurate, high-frequency monitoring of the atmospheric opacity along the line-of-sight of the telescope and a measure of the instrument responsivity and transmission efficiency at the time of the observation. The methods used to calibrate SCUBA-2 data were based primarily on the successful methods of calibration used for SCUBA, and details of these methods and their results can be seen in Archibald et al.\cite{archibald}, Jenness et al.\cite{jenness2} and Stevens $\&$ Robson\cite{stevens}. Commissioning and the shared-risk observing  carried out in December 2009 to March 2010 included a program of skydips to monitor the atmospheric opacity and stability and to investigate the transmission properties of the SCUBA-2  450$\mu$m and 850$\mu$m bandpass filters. Nightly observations of astronomical sources with known flux properties were used to determine the instrument performance and to provide absolute flux calibration of the instrument. This paper presents the preliminary results from these commissioning observations.\\

\section{EXTINCTION CORRECTION AT SUBMILLIMETER WAVELENGTHS}

\subsection{Atmospheric attenuation} 
\label{sec:atm}

\begin{figure}[h]
   \begin{center}
   \begin{tabular}{c}
   \includegraphics[height=7cm,width=12cm]{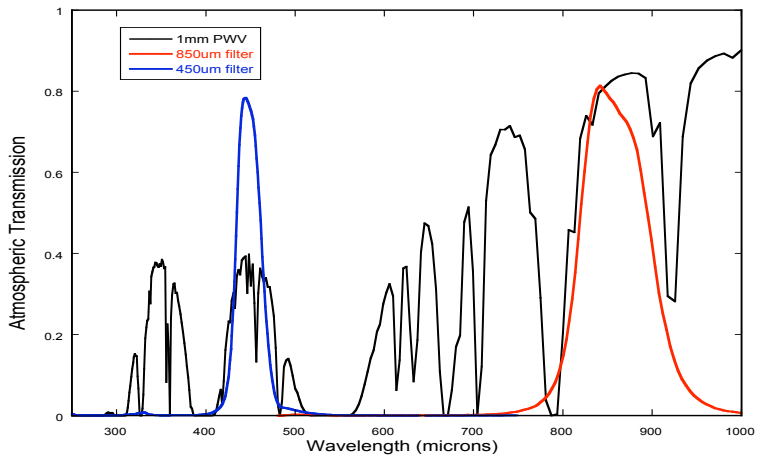}
   \end{tabular}
   \end{center}
   \caption[example]
   { \label{fig:atm}
450$\mu$m  (blue) and 850$\mu$m (red) bandpass filters superimposed on the submillimeter atmospheric transmission curve for Mauna Kea assuming 1mm of precipitable water vapour.}
   \end{figure}

The atmosphere severely limits ground-based observations at submillimeter wavelengths, providing only a few semi-transparent windows even at a high, dry site such at the summit of Mauna Kea, Hawaii. SCUBA-2, like its predecessor, SCUBA (Holland et al.\cite{holland}), has been designed to take advantage of the 850$\mu$m and 450$\mu$m atmospheric windows. The SCUBA-2 bandpasses at both wavelengths are shown superimposed on the submillimeter atmospheric transmission windows in Figure~\ref{fig:atm}. The opacity in the submillimeter is primarily caused by water vapour, although both oxygen and ozone contribute. The transmission is wavelength dependent, with the shorter wavelength windows having lower transmission and faster susceptibility to worsening conditions. The water vapour column, and therefore the opacity of the atmosphere, can change on short timescales, and must be monitored at high frequency in order to correct for these effects.\\

Opacity conditions for night-time observing at the JCMT are assessed via the adjacent CSO 225GHz radiometer. The radiometer measures the sky opacity using a multiple-point skydip once every fifteen minutes. The JCMT has a 183GHz water vapour monitor (WVM) installed in the receiver cabin which measures the precipitable water vapour (PWV) along the line of sight at one second intervals\cite{wiedner1} . The PWV measured by the WVM is scaled with an empirical correction to produce a CSO-equivalent 225GHz opacity. The WVM and CSO values show excellent correlation for the stable part of the night between 9pm and 3am, with deviations occurring in the early evening and at sunrise when conditions become more variable. The $\tau_{wvm}$ to $\tau_{cso}$ relation is shown to be~\cite{dempsey} : 
\begin{equation}
\tau_{wvm} = 1.01 * \tau_{cso} + 0.01
\end{equation}

Previous investigations of the submillimeter opacity effects on bolometric observations at the JCMT~\cite{archibald} , reference the CSO opacity. This work will use the WVM measurements, using the above relation, as the WVM provides the time resolution necessary to analyse the high-frequency opacity changes that can significantly affect SCUBA-2 observations.\\

\subsection{Skydips}
\label{sec:skydips}

The standard method for estimating the zenith sky opacity at a given wavelength and azimuth is by performing a skydip. SCUBA measured the sky brightness temperature directly as a function of elevation, with absolute temperature calibration achieved by observing a hot and cold load of known temperature\cite{holland}. The SCUBA data was then fitted to a model describing both a plane-parallel atmosphere and the optical system\cite{archibald}:

\begin{equation}
J_{meas} = (1 - \eta_{tel}) J_{tel} + \eta_{tel} J_{atm} - bwf~\eta_{tel}~J_{atm}~exp( - Tau * A)
\end{equation}

\noindent
where $J_{meas}$ is the measured sky brightness, $J_{tel}$ and $J_{atm}$ are the brightness temperatures of the telescope and atmosphere, respectively and $\eta_{tel}$ is the transmission of the telescope. The final variable, $bwf$, is the bandwidth factor of the filter being used, and is a function of water vapour content. The sky temperature is not measured directly by SCUBA-2. However, the detectors measures the change in optical loading with high accuracy. By measuring the change in optical load when opening a 1 K cold shutter located in front of the detector arrays an accurate measure of the optical load from the optics and sky is obtained. We fit this power change, $\Delta$P$_{heat}$ to the model. A full calibration of the relation between sky temperature and $\Delta$P$_{heat}$ was not completed during the initial commissioning run, however the response between the two is a linear relation. The data can therefore be fitted using a function of the form:

\begin{equation}
\label{eq1}
\Delta P_{heat} = P_0 - P_1 ~exp ( -(\alpha * (\tau_{wvm} - \beta)) * A)
\end{equation}

\begin{figure}[h]
   \begin{center}
   \begin{tabular}{c}
   \includegraphics[height=7cm]{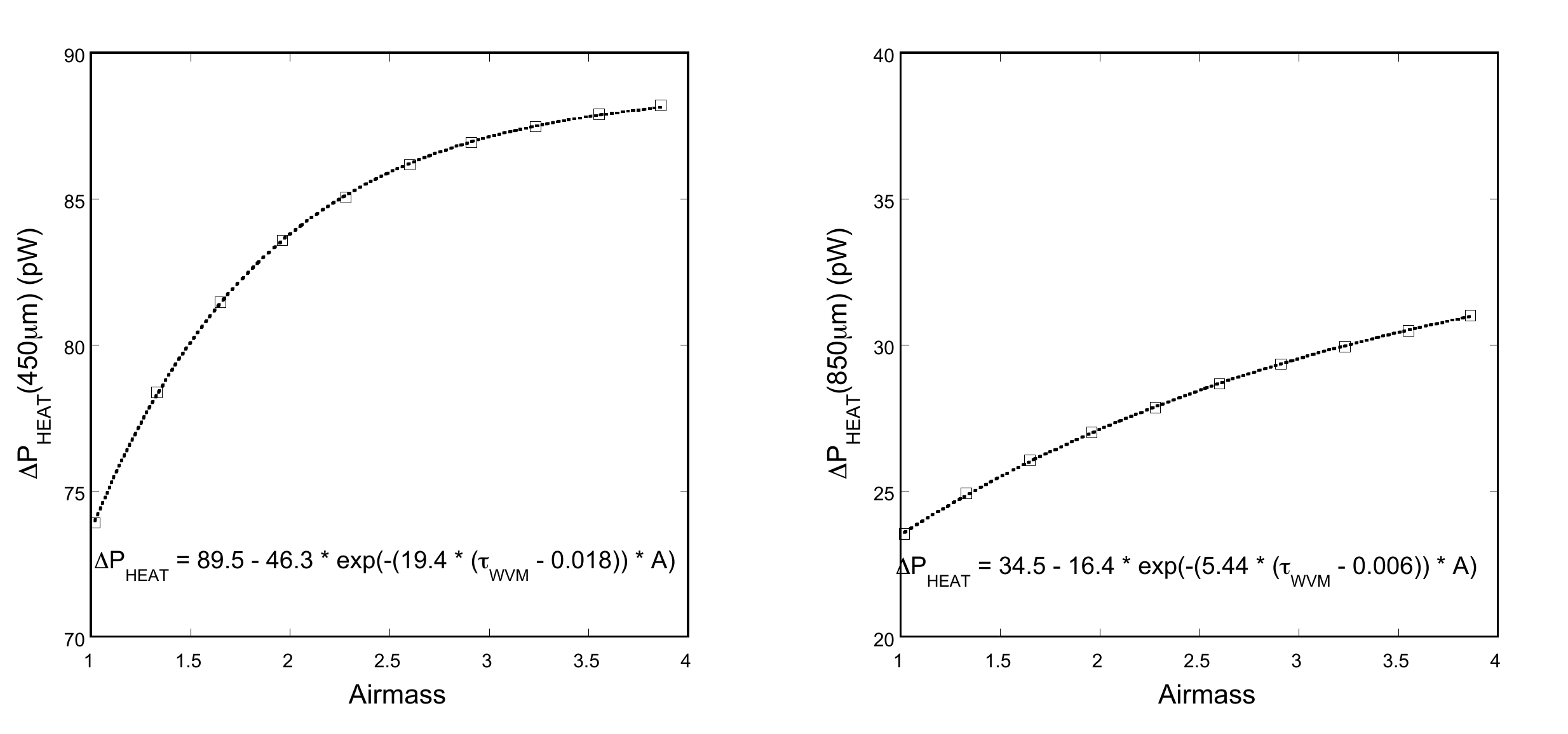}
   \end{tabular}
   \end{center}
   \caption[example]
   { \label{fig:dips}
450$\mu$m (left) and 850$\mu$m (right) discrete skydip examples taken simultaneously at a PWV $\approx$ 1mm. Square points are the heater power change $\Delta$P$_{heat}$, as described in section~\ref{sec:atm}, at each airmass, and the dashed line is the model fit to the data.}
   \end{figure}

This simple model does not take variation in temperature of the atmosphere but is accurate enough for the current purpose. A skydip is obtained by measuring the change in optical loading at a number of elevations. An example of fitting the equation to such a skydip is shown in Figure~\ref{fig:dips}.\\

\subsection{Tau relations}
\label{sec:tau}

Until sufficient calibration data could be collected on the sky with SCUBA-2, it was assumed that the SCUBA-2 filters as shown in Figure~\ref{fig:atm} were similar enough in their transmission properties to the SCUBA filters that the tau relations derived for SCUBA would be adequate for preliminary extinction correction. The original SCUBA opacity correction terms, as determined by Archibald et al.~\cite{archibald} were $\tau_{850}$ = 4.05($\tau_{wvm}$ - 0.001) and $\tau_{450}$ = 26.2($\tau_{wvm}$ - 0.014) .\\

Fifty skydips were completed in the period between February 22nd and March 14th, 2010. The individual dips were fitted to Equation~\ref{eq1}, and an example of an individual fit of model parameters are shown in Figure~\ref{fig:dips}. The advantage of the SCUBA-2 detector array design means that each time the shutter is opened from the fixed, closed shutter heater value, a new data point for $\Delta$P$_{heat}$ is obtained. Thus, every observation, not only skydips, can be used to determine the functional form of the relation between the sky power and the atmospheric transmission. If the wavelength-dependent $\tau$ is accounted for correctly, the power to transmission relation should be linear. With the original SCUBA opacity corrections, the power-transmission data showed distinct curvature, indicating that these relations did not correctly account for the opacity as seen by the individual arrays. When the new relations were fitted to the skydips and then used to calculate the atmospheric transmission, of the form Transmission = exp(-$\tau\times$A), where $\tau$ is the opacity and A is the airmass, the data showed an excellent linear correlation at both wavelengths.\\

\begin{figure}
   \begin{center}
   \begin{tabular}{c}
   \includegraphics[height=7cm]{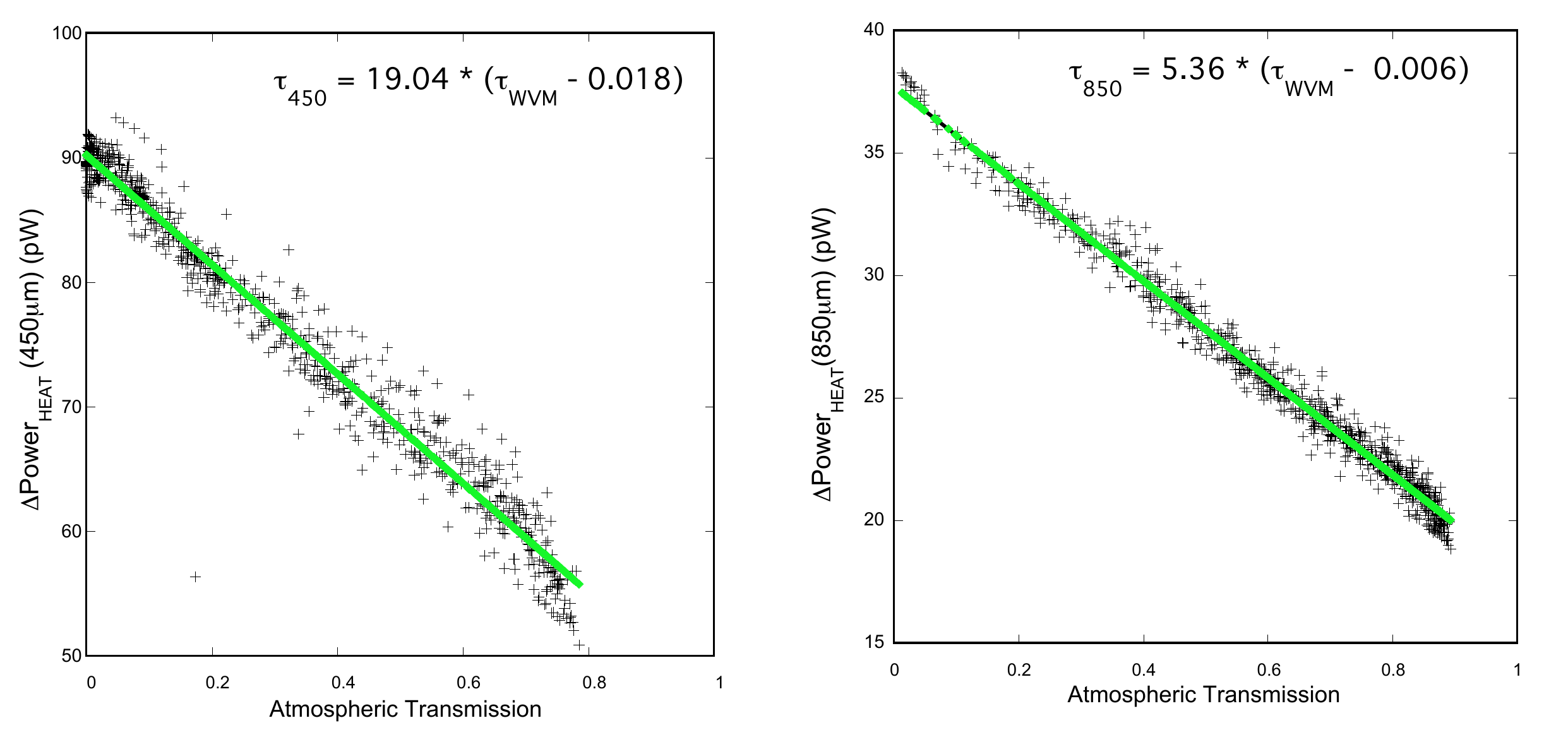}
   \end{tabular}
   \end{center}
   \caption[example]
   { \label{fig:heat}
The crosses denote the 450$\mu$m (left) and 850$\mu$m (right) $\Delta$P$_{heat}$ for all observations between the 22nd of February and 14th of March 2010, as a function of atmospheric transmission, overlaid with the model fits to the skydip data (squares).}
   \end{figure}

Figure~\ref{fig:heat} displays the results of the fit to Equation~\ref{eq1}. The crosses in both figures indicate all $\Delta$P$_{heat}$ for all observations taken between the 22nd of February and the 14th of March, 2010. Overlaid as green squares are the fitted data points from the fifty skydips obtained in that period. The final fitted form of the tau relations for each wavelength are described as: $\tau_{850}$ = 5.36($\tau_{wvm}$ - 0.006) and $\tau_{450}$ = 19.04($\tau_{wvm}$ - 0.018). 

\subsection{Astronomical calibrators}
\label{sec:fcf}

Nightly flux calibration is achieved by observation of astronomical sources with known flux properties. Mars and Uranus were the primary calibrators, and were used predominantly in preliminary commissioning. During the observing period analysed here, the bulk of calibration observations were undertaken on a series of secondary calibrators, which are listed with their SCUBA fluxes in Table~\ref{tab1}.

\begin{table}[h]
\caption{Secondary calibrators used for flux calibration of SCUBA-2. The flux values are sourced from the references noted in the table. }
\label{tab1}
\begin{center}
\begin{tabular}{|c|c|c|c|c|} 

\hline
\rule[-1ex]{0pt}{3.5ex} Source & RA(J2000) & DEC(J2000) & 850$\mu$m flux & 450$\mu$m flux  \\
\hline
\rule[-1ex]{0pt}{3.5ex} HL Tau & 04 31 38.4 & +18 13 59.0 & 2.36 $\pm$ 0.24\cite{jenness2}  & 9.9 $\pm$ 2.0\cite{jenness2}\\
\hline
\rule[-1ex]{0pt}{3.5ex} CRL 618	& 04 42 53.60 & +36 06 53.7 & 4.7  $\pm$ 0.37\cite{jenness2}  & 12.1 $\pm$ 2.2\cite{jenness2} \\
\hline
\rule[-1ex]{0pt}{3.5ex} CRL 2688 & 21 02 18.81 & +36 41 37.7 & 6.39  $\pm$ 0.51\cite{jenness2}  & 30.9 $\pm$ 3.8\cite{jenness2} \\
\hline
\rule[-1ex]{0pt}{3.5ex} IRC + 10216 & 09 47 57.38 & +13 16 43.7 & 8.8  $\pm$ 1.1\cite{jenness2}   & 17.5$\pm$ 4.5\cite{jenness2} \\
\hline
\rule[-1ex]{0pt}{3.5ex} V883 Ori &  05 38 19  & -07 02 2.0 & 1.34 $\pm$ 0.01\cite{barnard}  & 7.28 $\pm$ 0.07\cite{barnard}   \\
\hline
\rule[-1ex]{0pt}{3.5ex} Alpha Ori & 5 55 10.31 & 07 24 25.4 & 0.628 $\pm$ 0.008\cite{barnard}  & 1.39 $\pm$ 0.04\cite{barnard}  \\
\hline
\rule[-1ex]{0pt}{3.5ex} TW Hydrae & 11 01 51.91 & -34 42 17.0 & 1.37 $\pm$ 0.01\cite{barnard}  & 3.9 $\pm$ 0.7\cite{barnard}  \\
\hline
\rule[-1ex]{0pt}{3.5ex} Arp 220 & 15 34 57.21 & +23 30 09.5 & 0.668 $\pm$ 0.007\cite{barnard}  & 2.77 $\pm$ 0.06\cite{barnard} \\
\hline

\end{tabular}
\end{center}
\end{table}

When an observation of a source is run through the map-making software it has been flat-fielded and extinction corrected, and the resulting signal is in picowatts (pW). To convert the measured signal of a source from pW to Janskys, a `flux conversion factor' (FCF) needs to be applied to the map. These FCFs are determined by regular observation of planets, when available, and the sources listed in Table~\ref{tab1}. It is ideal to observe calibrators at least a  few times over each night, to account for any variation in instrument performance and dish temperature changes over the course of the evenings observing.\\

\begin{figure}[h]
   \begin{center}
   \begin{tabular}{c}
   \includegraphics[height=7cm]{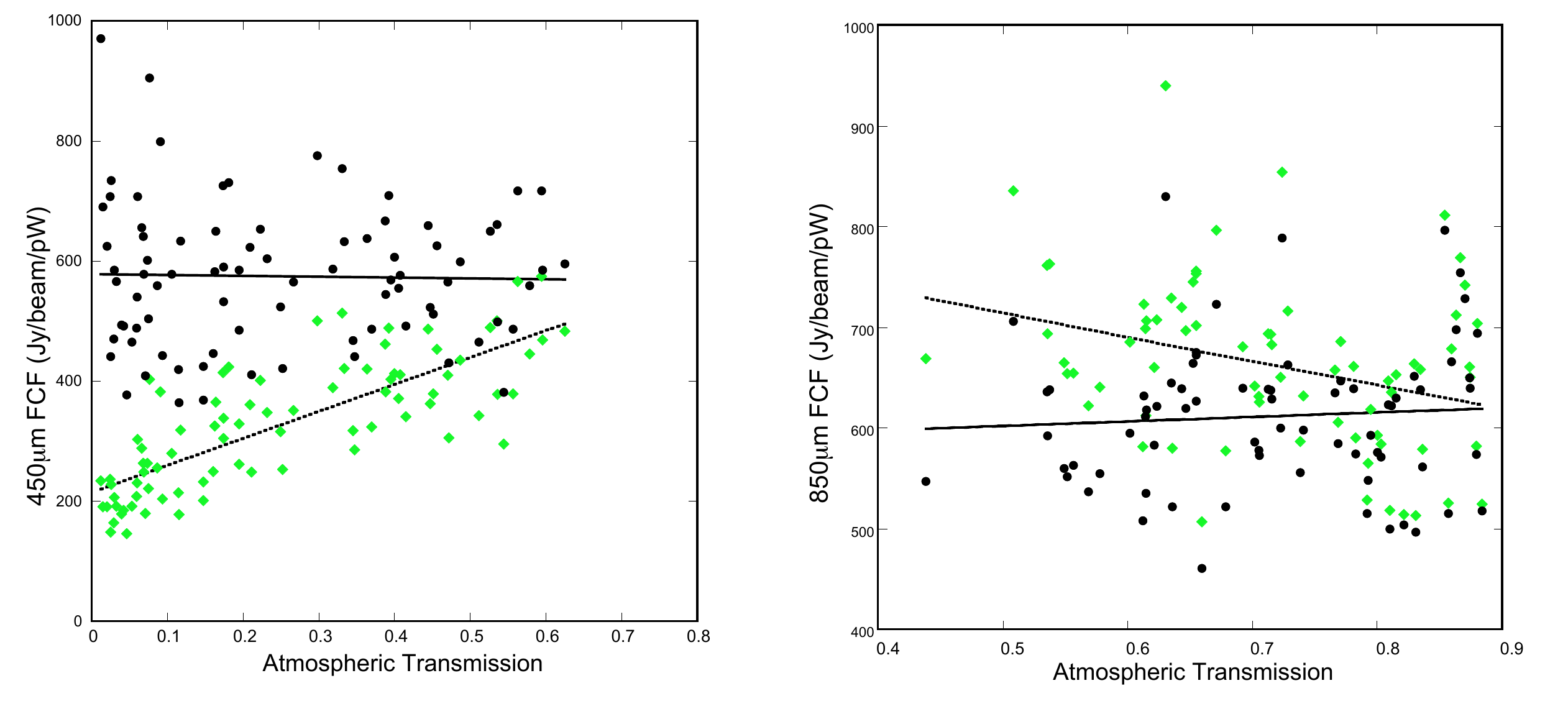}
   \end{tabular}
   \end{center}
   \caption[example]
   { \label{fig:fcf}
The 450$\mu$m (left)  and 850$\mu$m (right) flux conversion factors (FCFs), in Jy/beam/pW, for all secondary calibrators observed in the commissioning period, as a function of atmospheric transmission. In both cases, the green squares indicate the FCF calculated with the SCUBA tau relation, and the black circles plot the FCF as calculated with the tau relation fitted to the SCUBA-2 skydips. The linear fits show that in both cases, the new tau relation successfully removes a significant transmission dependent term in the FCF data.}
   \end{figure}

The FCF can be calculated in two ways. Firstly from the `peak' signal of a point source, such that $FCF_{peak} = S_{peak} / P_{peak}$ where $S_{peak}$ is the flux in the beam and $P_{peak}$ is the measured peak power (pW) from the map, giving units of Jy beam$^{-1}$. Secondly the integrated value over an aperture can be calculated in units of Jy arcsec$^{-2}$, where $FCF_{int} = S_{tot} / [P_{int} A]$.  $S_{tot}$ is the total flux of the calibrator, $P_{int}$ is the integrated sum in pW and A is the pixel area in arcsec$^2$. The method using the integrated flux was found to produce more consistent results as the peak value is susceptible to telescope effects such as focus and pointing drifts. Therefore, a third method was used for SCUBA-2 where the $FCF_{int}$ was taken and modeled with a Gaussian beam with a FWHM equivalent to that of the JCMT beam at each wavelength. The resulting FCF, denoted FCF$_{beamequiv}$, calculates a peak FCF from the integrated value assuming that the point source is a perfect Gaussian. The equation for FCF$_{beamequiv}$ is written as: 
\begin{equation}
\label{eq:fcf}
FCF_{beamequiv}  = \frac{S_{tot} \times 1.133 \times FWHM_{beam}^2}{P_{int} \times A}
\end{equation}

\noindent
where the FWHM$_{beam}$ for the 450$\mu$m is 7.25$''$ and the 850$\mu$m is 14.5$''$, respectively. Ideally, when the instrument is stable and the flat-fielding and extinction correction are correctly accounted for, the FCF should be a constant factor. However, when the secondary calibrator FCF$_{beamequiv}$ values were calculated for the observing period from February 22nd to the 14th of March, and were plotted as a function of transmission, a trend could be observed that was particularly obvious at 450$\mu$m. This is shown by the green squares in Figure~\ref{fig:fcf}. When the new tau conversion factors calculated here are reapplied to the FCF's, the trend as a function of transmission is removed at both wavelengths, indicating that the new opacity corrections are approximating the extinction correction more effectively.\\

\section{SUBMILLIMETER SEEING} 
\label{sec:seeing}

Submillimeter seeing is caused by fluctuations in the refractive index of the atmosphere, largely caused by the passage of water vapour passing through the beam. The variations caused by this inhomogeneity are of particular concern for interferometers such as the Submillimeter Array (SMA), adjacent to the JCMT. The Smithsonian Astrophysical Observatory (SAO) operates a phase monitor which tracks the submillimeter seeing by measuring the path length shifts induced by the water vapour fluctuations in the atmosphere. A description of the SAO phase monitor can be found in Masson (1994)~\cite{Masson}. The phase difference is measured each second and then one minute's worth of data are analysed for the rms scatter to provide the seeing at 12GHz over three baselines. \\

It was predicted that seeing would produce pointing offsets, however no correlation was seen between the SAO seeing and pointing excursions in SCUBA~\cite{archibald}.  Beam-broadening can occur if atmospheric turbulence on scales smaller than the telescope is significant. Experiments by Church and Hills\cite{church} suggested that this was unlikely to be a large effect at JCMT.  Attempts have been made previously to find a correlation between the seeing at the JCMT and the SMA seeing observed by the phase monitor\cite{archibald}. However, the standard SCUBA integration times were too long to allow for any direct measurement of the seeing to be reliably obtained, as seeing fluctuations instead occur on the order of a second or less. SCUBA-2, with a sampling rate of 200Hz, has the capability to image at these timescales. A recent data set taken on a night of extremely low tau ($\tau_{wvm} \approx 0.03$) yet extremely unstable localised weather prompted a close examination of the seeing using SCUBA-2 data.\\

\subsection{Mars} 
\label{sec:mars}

\begin{figure}
   \begin{center}
   \begin{tabular}{c}
   \includegraphics[width=15cm]{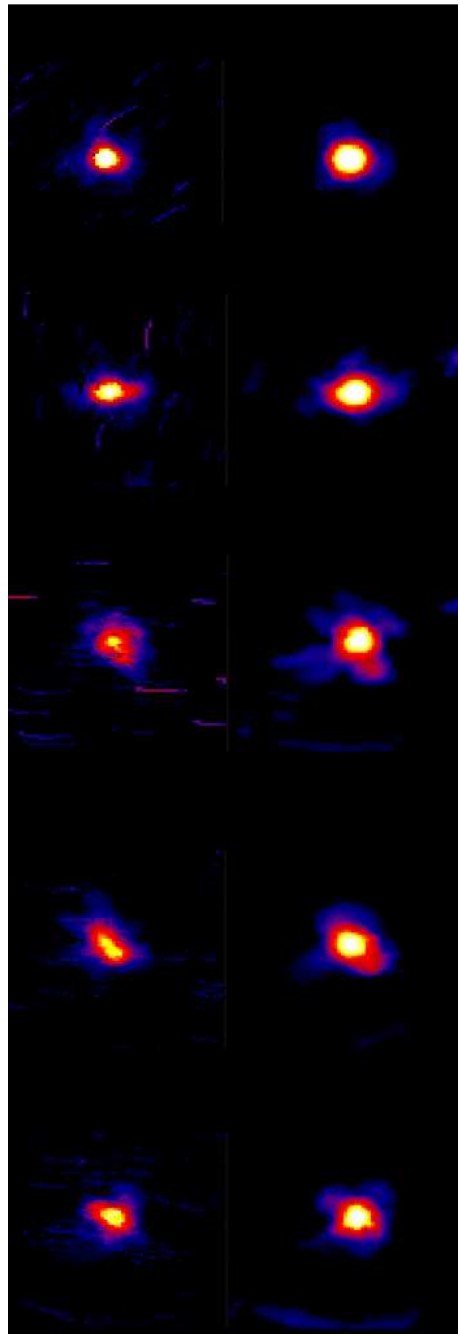}
   \end{tabular}
   \end{center}
   \caption[example]
   { \label{fig:seeing}
The 450$\mu$m (left) and 850$\mu$m (right) Mars image (0.4 second samples) over the course of a minute's observation. Time progresses from top to bottom.}
   \end{figure}

The Mars observation shown in Figure~\ref{fig:seeing} was obtained on the evening of March 10th, 2010. Early evening faults included a high rate of sky flat-field failures and inconsistent pointing results. The tau, though low, was extremely unstable. A mode of the SCUBA-2 reduction software allows the reduction of small time-slices of sequential data to be made into individual maps. With a source as bright as Mars, these maps can be made of only a few tenths of a second worth of data, and still have sufficient signal-to-noise for analysis. The Mars observation was a minute long, and individual maps span 0.4 seconds of data each.  Figure~\ref{fig:seeing} is a series of snapshots of these data over a minute, progressing in time from top to bottom. \\

\begin{figure}
   \begin{center}
   \begin{tabular}{c}
   \includegraphics[height=7cm]{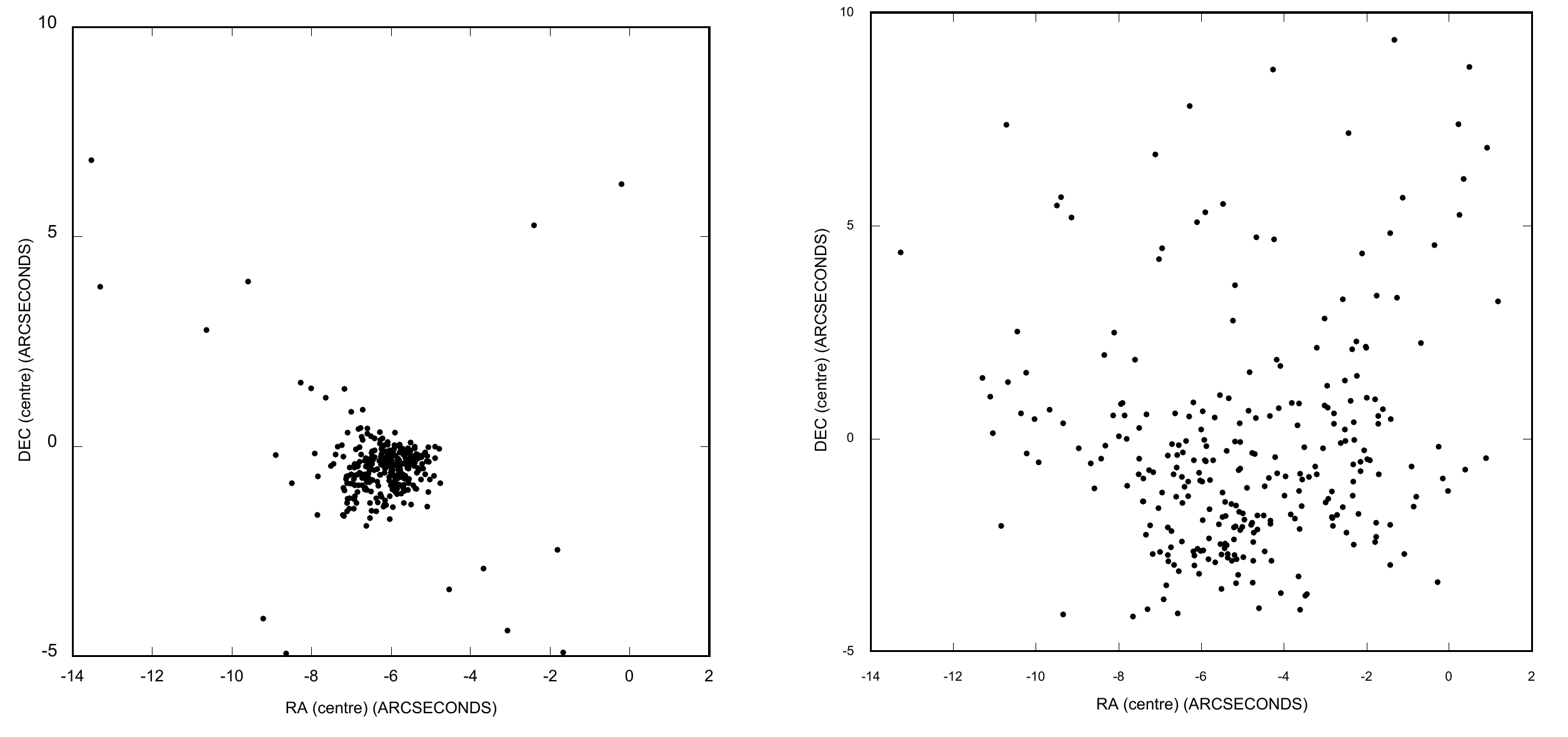}
   \end{tabular}
   \end{center}
   \caption[example]
   { \label{fig:seeing2}
The 450$\mu$m offset centroid positions in RA and DEC (arcseconds) of Mars in 0.4 second slices for one-minute observations on the 24th of February (left) and the 10th of March (right).}
   \end{figure}

As can be seen from the image, both pointing shifts and beam distortion occur, and both effects can be seen to correlate between the two wavelengths. To determine the magnitude of the distortion caused by the seeing, the individual maps were analysed using the KAPPA\cite{kappa} beamfit program. Figure~\ref{fig:seeing2} shows the measured centroid positions of Mars in each 0.4 second time-slice for a night (February 24th) of stable conditions ($\tau_{wvm} = 0.07$) and March 10th ($\tau_{wvm} = 0.03$). The left hand plot is typical of Mars observations taken throughout this time period. \\

\begin{figure}
   \begin{center}
   \begin{tabular}{c}
   \includegraphics[height=7cm]{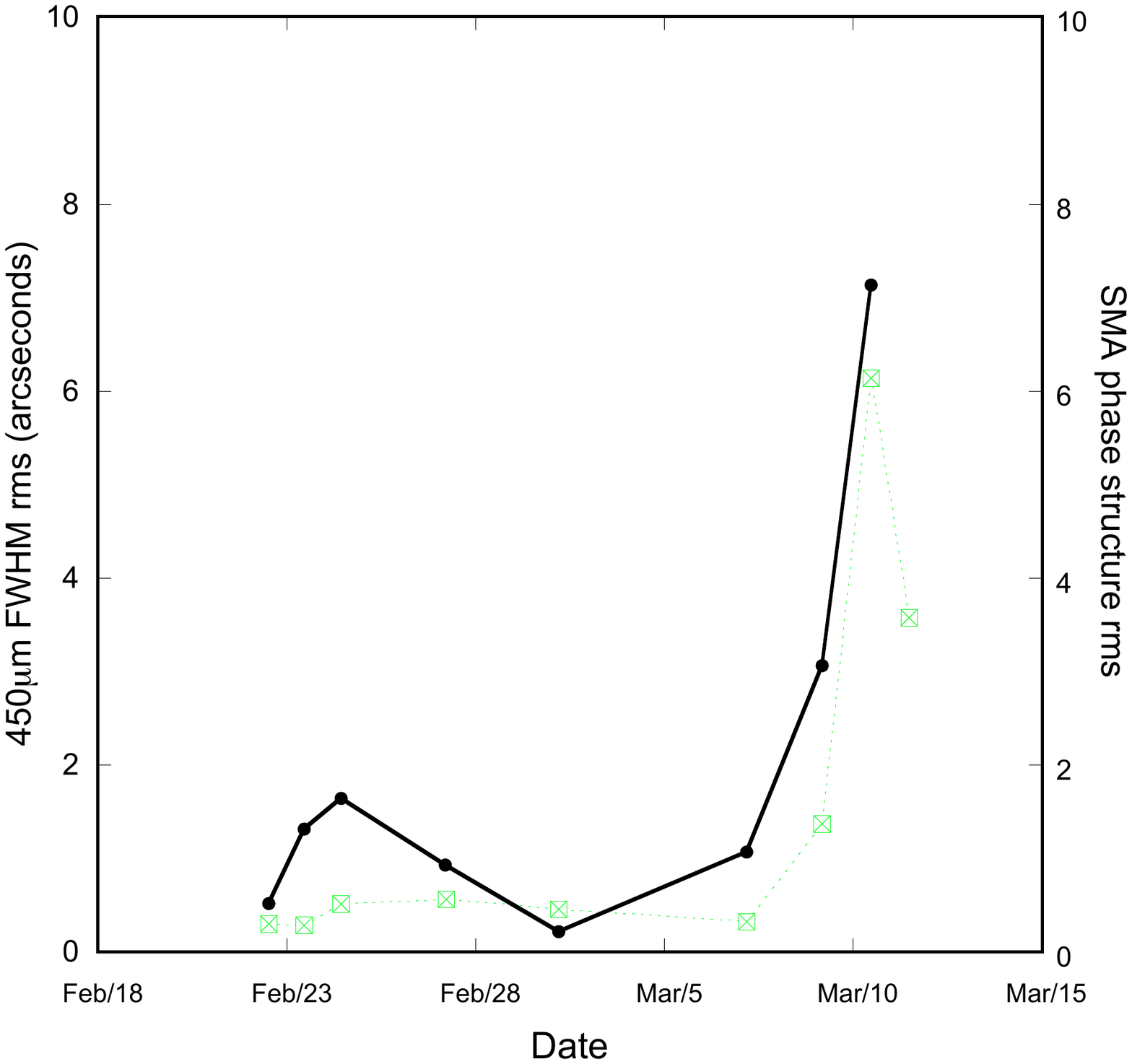}
   \end{tabular}
   \end{center}
   \caption[example]
   { \label{fig:seeing3}
The rms (arcseconds) of the 450$\mu$m FWHM (Mars) centroids for a series of observations between the 22nd of February and the 10th of March (black circles) versus the rms of a single selected baseline of the SMA phase monitor for the same nights (green squares). }
   \end{figure}

 Eight SCUBA-2 observations of Mars taken between the 22nd of February and the 10th of March were analysed using the method described above. The black circles and line in Figure~\ref{fig:seeing3} shows the rms of the 450$\mu$m FWHM of Mars in arcseconds. The green squares denote the rms phase of a single baseline of the SMA phase monitor for the same nights. The correlation between the SMA seeing and that observed with SCUBA-2 needs a more detailed analysis. The simple comparison in Figure~\ref{fig:seeing3} uses an average value for the SMA rms, calculated for the hour bracketing the time when the SCUBA-2 observation was taken. The trends in the two datasets show good agreement as the seeing deteriorates on the 9th and 10th of March.\\

\section{CONCLUSIONS} 
\label{sec:conc}

SCUBA-2 calibration results from the commissioning and shared-risk observing period between February 22nd and March 14th of 2010 are presented and analysed. The atmospheric opacity relations for the 850$\mu$m and 450$\mu$m bandpass filters are shown to be $\tau_{850}$ = 5.36($\tau_{wvm}$ - 0.006) and $\tau_{450}$ = 19.04($\tau_{wvm}$ - 0.018), respectively. These relations were determined by an empirical fit to the skydip data obtained during this period. The relations differ from those previously determined for the SCUBA instrument, which were being used in initial reductions of the science data taken with SCUBA-2. The flux conversion factors (FCF's) determined from observation of calibration sources were shown to have a trend as a function of atmospheric transmission at both wavelengths. Application of the new opacity terms in the extinction correction model removed this transmission dependence in the FCFs, thus improving the accuracy of SCUBA-2 calibration.\\

An observation of Mars on the 10th of March 2010 was shown to exhibit the effects of submillimeter seeing. SCUBA-2's high sampling rate allowed time-slices of 0.4 seconds in the data to be analysed, and these sequential images showed evidence of both beam distortion and pointing shifts which were shown to lie well outside of the normal range when compared to similar observations on previous nights.\\

\acknowledgments     
 
The authors acknowledge the support of the astronomy research funding agencies of the UK, Canada and The Netherlands. This work has used seeing data from the Submillimeter Array phase monitor operated by the Smithsonian Astrophysical Observatory, to whom we are grateful for providing these data.


\bibliography{spie2010}   

\begin{thebibliography}{10}

\bibitem{holland2}
{Holland}, W. S. e.~a., ``{SCUBA-2: first results and on-sky performance},'' in
  [{\em Society of Photo-Optical Instrumentation Engineers (SPIE) Conference
  Series}{\nolinebreak\hspace{0.1em}]},  {\em Society of Photo-Optical
  Instrumentation Engineers (SPIE) Conference Series} {\bf 7741},  4+ (Aug.
  2010).

\bibitem{bintley}
{Bintley}, D. e.~a., ``{Characterising the SCUBA-2 superconducting bolometer
  arrays},'' in [{\em Society of Photo-Optical Instrumentation Engineers (SPIE)
  Conference Series}{\nolinebreak\hspace{0.1em}]},  {\em Society of
  Photo-Optical Instrumentation Engineers (SPIE) Conference Series} {\bf 7741},
   5+ (Aug. 2010).

\bibitem{archibald}
{Archibald}, E.~N., {Jenness}, T., {Holland}, W.~S., {Coulson}, I.~M.,
  {Jessop}, N.~E., {Stevens}, J.~A., {Robson}, E.~I., {Tilanus}, R.~P.~J.,
  {Duncan}, W.~D., and {Lightfoot}, J.~F., ``{On the atmospheric limitations of
  ground-based submillimetre astronomy using array receivers},'' {\em
  MNRAS}~{\bf 336},  1--13 (Oct. 2002).

\bibitem{jenness2}
{Jenness}, T., {Stevens}, J.~A., {Archibald}, E.~N., {Economou}, F., {Jessop},
  N.~E., and {Robson}, E.~I., ``{Towards the automated reduction and
  calibration of SCUBA data from the James Clerk Maxwell Telescope},'' {\em
  MNRAS}~{\bf 336},  14--21 (Oct. 2002).

\bibitem{stevens}
{Stevens}, J.~A. and {Robson}, E.~I., ``{On Improving the Calibration of
  Millimetre and Submillimetre Photometry at the James Clerk Maxwell Telescope
  / JCMT},'' {\em MNRAS}~{\bf 270},  L75+ (Oct. 1994).

\bibitem{holland}
{Holland}, W.~S., {Robson}, E.~I., {Gear}, W.~K., {Cunningham}, C.~R.,
  {Lightfoot}, J.~F., {Jenness}, T., {Ivison}, R.~J., {Stevens}, J.~A., {Ade},
  P.~A.~R., {Griffin}, M.~J., {Duncan}, W.~D., {Murphy}, J.~A., and {Naylor},
  D.~A., ``{SCUBA: a common-user submillimetre camera operating on the James
  Clerk Maxwell Telescope},'' {\em MNRAS}~{\bf 303},  659--672 (Mar. 1999).

\bibitem{wiedner1}
{Wiedner}, M.~C., {Hills}, R.~E., {Carlstrom}, J.~E., and {Lay}, O.~P.,
  ``{Interferometric Phase Correction Using 183 GHZGHz Water Vapor Monitors},''
  {\em ApJ}~{\bf 553},  1036--1041 (June 2001).

\bibitem{dempsey}
{Dempsey}, J.~T. and {Friberg}, P., ``{Optimizing atmospheric correction at the
  JCMT using a 183GHz water vapor radiometer},'' in [{\em Society of
  Photo-Optical Instrumentation Engineers (SPIE) Conference
  Series}{\nolinebreak\hspace{0.1em}]},  {\em Society of Photo-Optical
  Instrumentation Engineers (SPIE) Conference Series} {\bf 7012},  70123Z--1 --
  70123Z--11 (Aug. 2008).

\bibitem{barnard}
{Barnard}, V.~E., ``{A summary of the search for new Submillimetre Secondary
  Calibrators},'' Tech. Rep. {SCD/SN/011}, {Joint Astronomy Centre} (2005).

\bibitem{Masson}
{Masson}, C.~R., ``{Atmospheric Effects and Calibrations},'' in [{\em IAU
  Colloq. 140: Astronomy with Millimeter and Submillimeter Wave
  Interferometry}{\nolinebreak\hspace{0.1em}]},  {Ishiguro}, M. and {Welch},
  J., eds., {\em Astronomical Society of the Pacific Conference Series} {\bf
  59},  87--+ (1994).

\bibitem{church}
{Church}, S. and {Hills}, R., ``{Measurements of daytime atmospheric 'seeing'
  on Mauna Kea made with the James Clerk Maxwell Telescope},'' in [{\em
  URSI/IAU Symposium on Radio Astronomical
  Seeing}{\nolinebreak\hspace{0.1em}]},  {J.~E.~Baldwin \& S.~Wang}, ed.,
  75--80 (1990).

\bibitem{kappa}
{Currie}, M.~J., {Draper}, P.~W., {Berry}, D.~S., {Jenness}, T., {Cavanagh},
  B., and {Economou}, F., ``{Starlink Software Developments},'' in [{\em
  Astronomical Data Analysis Software and Systems
  XVII}{\nolinebreak\hspace{0.1em}]},  {R.~W.~Argyle, P.~S.~Bunclark, \&
  J.~R.~Lewis}, ed., {\em Astronomical Society of the Pacific Conference
  Series} {\bf 394},  650--+ (Aug. 2008).

\end{thebibliography}
\bibliographystyle{spiebib}   

\end{document}